\documentclass[%
 reprint,
 amsmath,amssymb,
 aps,
]{revtex4-2}

\usepackage{graphicx}
\usepackage{dcolumn}
\usepackage{bm}
\usepackage{textgreek}
\usepackage{multirow}
 \usepackage{url}
\usepackage[utf8]{inputenc}
\usepackage{hyperref}

\begin{document}

\preprint{APS/123-QED}

\title{Pore Network and Medial Axis Simultaneous Extraction through Maximal Ball Algorithm}

\author{Mariane Barsi-Andreeta*$^{a}$,Everton Lucas-Oliveira$^{a}$,Arthur Gustavo de Araujo-Ferreira$^{a}$,Willian Andrighetto Trevizan$^{b}$,Tito José Bonagamba$^{a}$\\
}

\affiliation{
 $^{a}$São Carlos Institute of Physics${,}$University of São Paulo${,}$ PO Box 369, 13560-970 ${,}$ São Carlos${,}$ SP${,}$ Brazil,\\
 $^{b}$CENPES${,}$ Petrobras${,}$ Rio de Janeiro${,}$RJ${,}$ Brazil,\\
 *mariane.andreeta@usp.br
}%

\date{\today}

\begin{abstract}
Network Extraction algorithms from X-ray microcomputed tomography have become a routine method to obtain pore connectivity and pore morphology information from porous media. The main approaches for this extraction are either based on Max Ball Algorithm or Centerline/Medial Axis Extraction. The first is a robust method to separate the pore space into pore and throats, which provides the pore and throat sizes distributions directly. The second simplifies the medium by thinning the volume into a one voxel wide centerline that preserves the volume’s topological information. Since each method has drastically different implementations, it is not usual to use both to characterize the porous structure. This work presents a method to extract a simplified centerline of porous materials by adding few more steps to the well-established Max Ball algorithm. Results for sandstones and carbonate rock samples show that this Medial-Axis network can be used for single phase flow simulations, as well as preserves the mediums morphology.
\end{abstract}

\maketitle


\section{INTRODUCTION}

Porous space modelling began with the analogy between flow in porous media and a random resistor network, idealized by Fatt in 1956, on the endeavor to understand the transport properties of porous materials\cite{Fatt1956,XIONG2016101}. Since then, this area of research has greatly advanced with the improvement of imaging techniques, which made it possible to recover the morphology of porous media on Benchtop X-ray tomography scanners with resolution higher than 1 \textmu m \cite{Vaz2014,CNUDDE20131}. The higher resolutions also brought a need for methods to extract information from the huge amount of data. This lead to the development and application of Pore Networks Extraction algorithms not only in the modelling of the pore space of rock and sand, with applications on aquifer management and oil recovery, but also to other applications that emerged, such as the evaluation of cellular materials \cite{Benouali2005}, trabecular bone structure investigation \cite{Mondal2019},the assessment of pore connectivity on lithium battery components \cite{Lagadec2018} and in tortuosity characterization of cement hydrates in high performance concrete \cite{Song2019}. These applications use Pore Network modelling not only to recover macroscopic parameters of transport simulations, but to do morphological analysis of the media.

An intuitive method to quantify the morphology of interconnecting channels is the simplification of the volumetric data through Medial Axis Extraction Algorithms (MA). These algorithms, based on thinning procedures, transforms the volumetric data into a one voxel wide skeleton that allows the quantification of the morphological parameters as channel tortuosity or diameter variation, preserving the medium's original topology. The quantification of these features have been used in various fields, but most commonly in medical applications, for example in the investigation of vascular disease \cite{Dougherty2000}. The main issue with current thinning algorithms is their sensitivity to the surface roughness on the boundary of the void space, therefore making them not well suited for more complex porous materials as rock cores. Even though new methods have been developed with alternatives to avoid this issue and preserve the topology of fractures \cite{Jiang2007,Jiang2017,Li2018}, the algorithms are usually not simple to implement and are still not very wide spread.

An alternative to the MA method is the Pore Network Extraction based on the separation of the pore space into two entities: pore bodies (nodes) and the connecting throats (edges). Since its development and further improvements, the Maximal-Ball Algorithm (MBA) has become a preferred method of this model of Pore Network extraction \cite{Silin2006,Blunt2016,Dong2009,Arand2017,Raeini2017}. It is very robust, able to retain the topology and morphology of different mediums, and the concepts behind the algorithm are easy to understand and implement. Examples include applications in shale \cite{Zheng2019}, gas bubble migration in sediment porous media \cite{Mahabadi2018}, and transport properties in sea shallow sediments \cite{Merey2019}.

The purpose of this work is to describe a method to simultaneously extract the medial axis from the volume as well as the pore-throat network just by adding a few strategic steps to the original MBA extraction. This providential modification in the algorithm provides an alternative to the approach of the medial axis, which, although having fewer nodes, can also capture the lengths and curvatures of the paths.

\begin{figure*}
\includegraphics[scale=0.4]{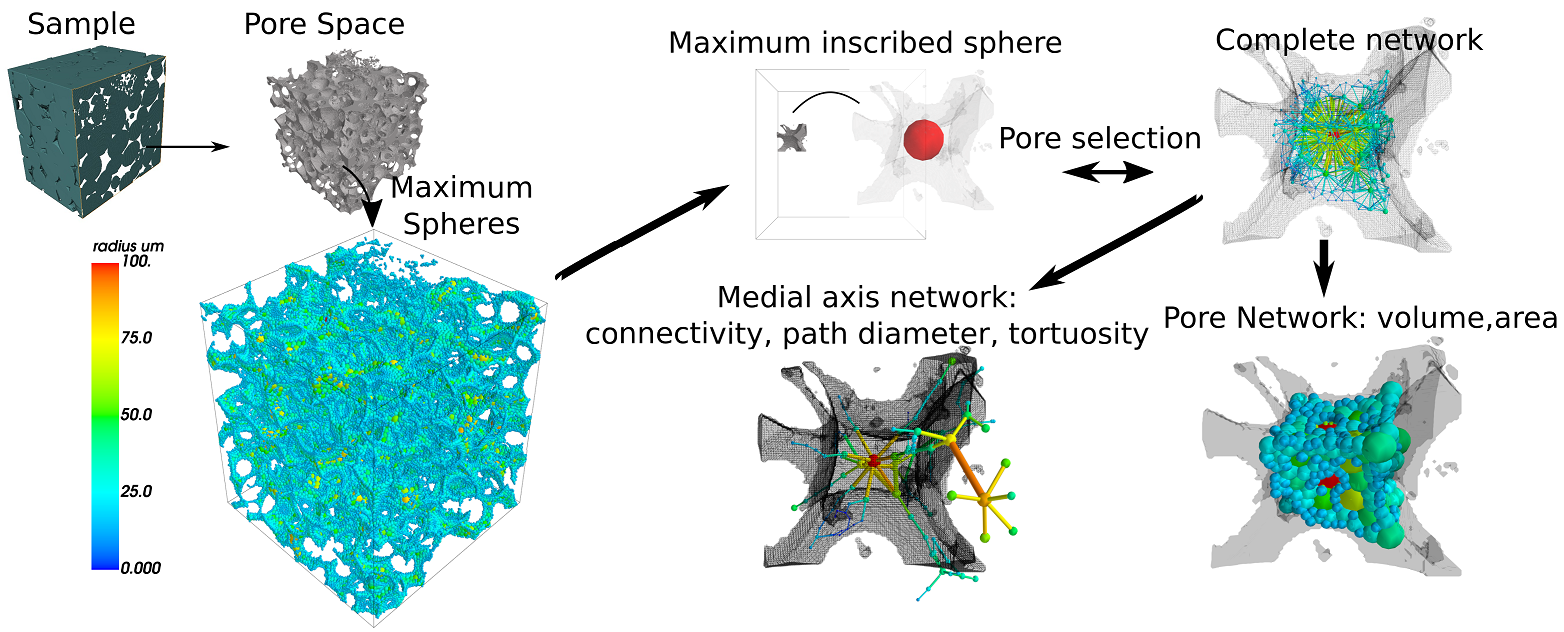}
\caption{\label{fig:summary}Graphical representation of the algorithm steps. First, the pore space is identified from the sample and a complete network of inscribed spheres is created. Then, the local maxima spheres are identified as pores and through Djikstra's algorithm, two outputs are generated: medial axis and pore-throat network.}
\end{figure*}

The idea is based on the pore-throat chains described on the original MBA algorithm \cite{Dong2009}. After the Maximal Balls are found and the Balls representing the pore centers and throat are defined, a pore-throat chain of spheres is built based on the algorithm process of hierarchy. The pore-throat chains are composed by the largest spheres that connect the pore spheres to the throat spheres. By connecting the chains, the final paths will compose a network that is a close approximation to the Medial-Axis network of MA algorithms.  The network will preserve the medium’s topology, by maintaining the original connections and represent the center of the local pore space preserving the medium's morphology. 

The main issue with the construction of the pore-throat chains in the MBA is dealing with groups Maximal Balls of same or very similar radii. On the original MBA, this is overcome by a clustering process. Here a new method is proposed solely based on Djikstras' algorithm for finding shortest paths in weighted graphs \cite{Dijkstra1959,KNUTH19771,coronary19}. The process is based on finding optimum paths between pores by modelling the Maximal Balls as a large weighted network. Djikstras' algorithm is a well-established algorithm with implementation available in most graph processing packages, being a simple and robust solution for the medial network extraction. 
The algorithm was tested for different weighting approaches for samples of carbonate and sandstones. The Medial-Axis networks were verified for the ability to predict the samples absolute permeability and the preservation of its morphology. After the Medial-Axis networks were extracted, the networks were further processed to acquire the  Pore-Throat Networks. 
Since the method is an extension of the MBA, a quick review of the basis of the original algorithm is presented followed by the proposed alterations. For further reading of the original method, the reader should refer to the cited previous works \cite{Blunt2016,Dong2009,Arand2017,Al-Kharusi2007,Silin2006}.

\section{ALGORITHM DESCRIPTION}

The set of voxels in the 3D matrix representation of the porous medium can be divided into two subsets:V\textsubscript{void} and V\textsubscript{material}. The first represents the entire pore space and the second the solid space of the material. The original MBA performs a search on each of the v\textsubscript{void}\textsuperscript{i} $\in$ V\textsubscript{void} voxels and calculates the maximum inscribed sphere S\textsuperscript{i}  centered at v\textsubscript{void}\textsuperscript{i}(x,y,z). A sphere is maximum if it is wholly contained in the void space, touches the material surface at least at one point and it is not included in other larger spheres.  

Next, the spheres are sorted according to their radius, from largest to smallest. Each sphere's radius is checked and compared to that of its neighbors. If it is the largest of the region it defines a Pore Center. Other spheres that have a connection to the Pore Center Sphere become part of the pore's body. The spheres that are assigned to two or more pores in this process are defined as throats. This definition is achieved through the algorithm's hierarchy process.

A summary of the basic steps is as follows:

\begin{enumerate}
    \item	Pore space binarization and definition;
    \item	The pore space is subdivided into maximal inscribed spheres. A sphere is said to be maximum if it is not contained by any other sphere and touches the volume surface at least at one point;
    \item	The spheres’ centers with the largest radius in each region is the center of a pore, and a tag is assigned to it;
    \item	Other spheres that touch a pore sphere, are assigned the same tag and are said to belong to the pore's body. This tag is passed through to neighboring spheres interactively. This is similar to a breadth-first search on a graph. Each sphere with smaller radius that is connected (overlaps another at some degree) to another with larger radius is given the same tag. This procedure is often compared to the pore’s “family name” being passed through to the smaller spheres that compose its volume.
    \item	If an already tagged sphere is found by one with a different ‘family name’, it is tagged as a throat, or connection between pores.
\end{enumerate}

\subsection{Medial-Axis Network Extraction }

The mapping procedure makes use of the maximal spheres acquired through the MBA algorithm to build a large weighted network, in which the spheres’ centers are the nodes’ positions. In the developed method, the maximum spheres are found by first calculating the Euclidean Distance Transform (EDT) on the volumetric data. The EDT attributes to each voxel v\textsubscript{void}\textsuperscript{i} the minimum distance between v\textsubscript{void}\textsuperscript{i} and all v\textsubscript{material}\textsuperscript{k} $\in$ V\textsubscript{material}: 
\begin{eqnarray}
EDT(v_{void}^i)=min_{0<k<N_{material}}(\|v_{void}^i-v_{material}^k\|)
\label{eq:edt}.
\end{eqnarray}

\begin{figure}
\includegraphics[scale=0.4]{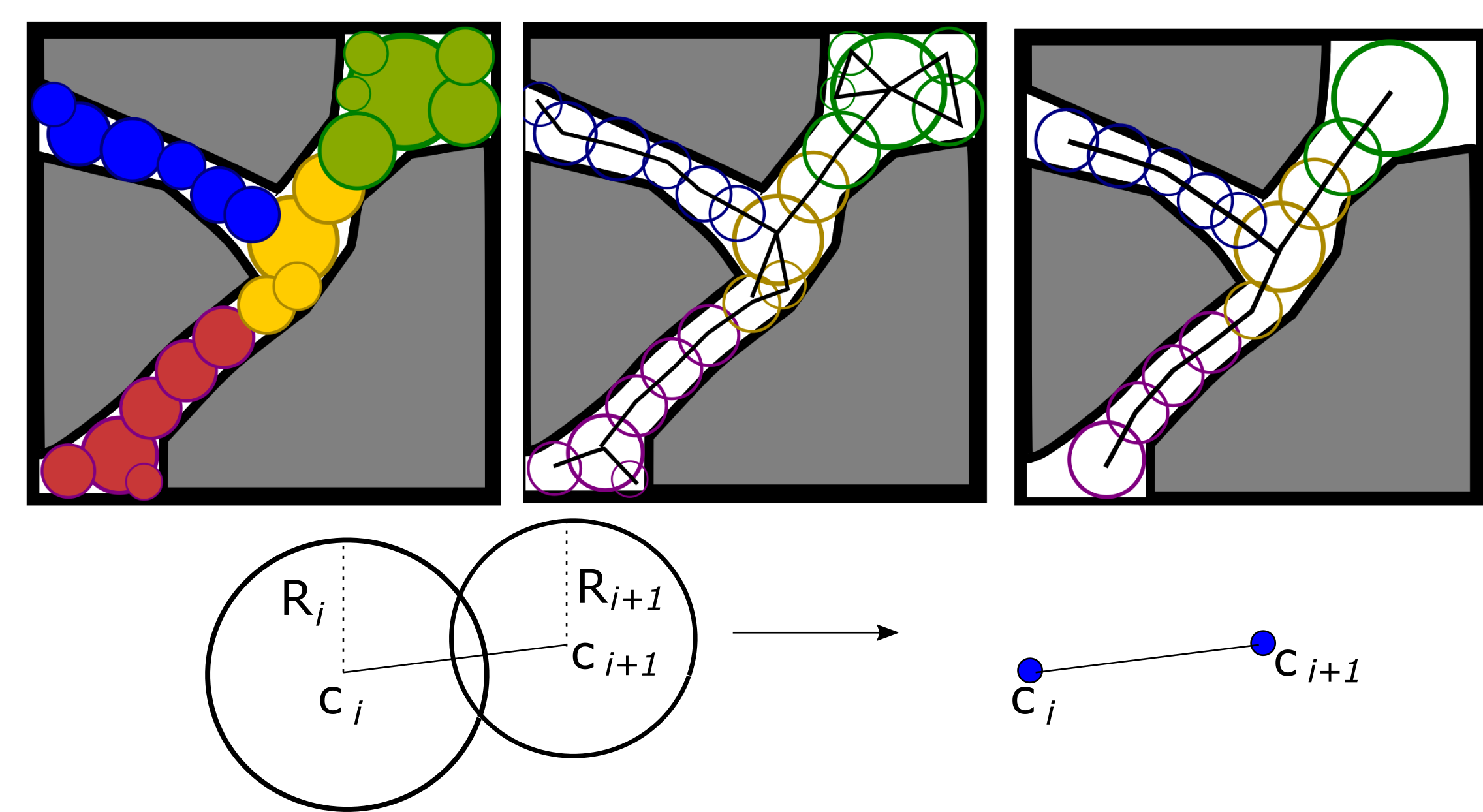}
\caption{\label{fig:graph01} The construction of the network of regions. Two separate networks are constructed: a network composed of the complete set of spheres (middle image) and a second one composed only of the medial axis spheres (right image). Each  node represents a sphere, stores its position,radius,effective volume and effective surface. Each sphere connection to its neighbor is weighted according to the proposed models.}
\end{figure}

The value attributed to each voxel by the EDT is the radius of the maximum inscribed sphere S\textsuperscript{i} at voxel  v\textsubscript{void}\textsuperscript{i} (x,y,z). Therefore, at this point, all possible inscribed spheres are “stored” in the volumetric data. The next step is to exclude spheres that are included in larger ones. A sphere S\textsuperscript{a} is included in a S\textsuperscript{b} with larger radius R\textsuperscript{b}, if 
\begin{eqnarray}
\|v_{a}^i-v_{b}^k\| <R^b
\label{eq:inclusion}.
\end{eqnarray}
A simple approach to directly avoid inclusions is to automatically exclude all voxels that belong to the current sphere's volume. A voxel v\textsubscript{void}\textsuperscript{j} (x,y,z) belongs to the sphere S\textsuperscript{i} with center coordinates c\textsubscript{void}\textsuperscript{i} (x,y,z) if:
\begin{eqnarray}
\|v(x,y,z)^j-c(x,y,z)^i\| <R^i
\label{eq:exclusion}.
\end{eqnarray}
The exclusion step tags with the S\textsuperscript{i}'s id all voxels v\textsubscript{void}\textsuperscript{j} that follow the rule of ~\ref{eq:exclusion}, if no tag were previously assigned to it. This step removes redundant information from the data set, and stores only the maximal spheres' information. It is important to note that equation ~\ref{eq:inclusion} allows for spheres to overlap to some degree. This fact is used to find the connections between spheres and build the network(Fig.~\ref{fig:graph01}).   

Each node stores its id; position; the sphere’s radius; effective volume; and effective contact area. The effective volume is the number of voxels tagged with the sphere's id in the exclusion step. The effective contact area is the number of voxels that belong to the effective volume set, but that have at least one neighboring material voxel v\textsubscript{material}.

After the sphere connections network is constructed, the pore nodes are defined by selecting the local maxima of the network defined by the spheres' radius, as described before on the original MBA. Then, the Medial Axis network is found by searching for the shortest path between the pore nodes in the complete network. 

The number of pore nodes used as seeds to find the optimum paths was left as a free parameter on the algorithm. The pore nodes are sorted from largest to smallest by their radii. Then a subset is selected to act as seeds from which the paths to all other pore nodes will be calculated. This improves the performance of the method and an evaluation of the impact of the selection of the pore nodes done in the previous step on the Medial-Axis extraction can be made.  

An additional parameter of the searching process is the cost set on each edge of the final network. The result of the search should be a single path between two pore nodes composed by largest possible spheres, with the least resistance for fluid flow. Therefore, an assumption is made that the Medial-Axis network will be well defined by using the spheres' radius as weighting strategy in the searching process. Three models were tested:
\begin{eqnarray}
Model 1: w(e_{(i,i+1)} )= 1
\label{eq:capacity01}.
\end{eqnarray} 
\begin{eqnarray}
Model 2: w(e_{(i,i+1)} )= 1/ R_{i}^2
\label{eq:capacity02}.
\end{eqnarray} 
\begin{eqnarray}
Model 3: w(e_{(i,i+1)} )= L_{i,i+1}/ R_{i}^4
\label{eq:capacity03}.
\end{eqnarray} 
The first model is to verify this assumption by defining the shortest path as the one with the least number of edges, with no weights applied. The second is based on the cross section area of the inscribed Maximal Ball, and the third is set based on Hagen-Pouseille law, in which L\textsubscript{i,i+1} is the distance between two spheres' centers and R\textsubscript{i} is set by selecting one of the spheres of the pair (Fig.~\ref{fig:graph01}). 

The extraction of the Pore-Networks followed the basic steps of the original hierarchy:
\begin{enumerate}
    \item	The spheres are sorted according to their radius (largest to smallest) and given infinite rank;
    \item	The spheres are divided according to size. Each group contains spheres of the same size;
    \item	The groups are processed from the largest size to smallest. Starting with the first sphere of the largest group, it passes its name to its smaller neighbors and ranks them one generation younger.  
    \item	The spheres left in the group are sorted according to their rank. If the next sphere's rank is infinite, it is given the tag of pore center, and sets its id as a family name. if not, it transfers the current family name to its smaller neighbors. 
    \item   The same sorting and clustering processes apply for all the groups 
    \item   If a sphere already belonging to a family is given a different family name, it is set as throat, and defines the connection between pores.
    \item   The process ends after processing all spheres.
\end{enumerate}

The diameter and length of the connection is set as twice the throat node's radius and as the Euclidean distance between the connecting pore centers respectively. 

This process is similar to the original MBA, but with some point differences. The original algorithm uses a pore merging method not applied here. Although the MBA algorithm is a robust approach to separate the pore space, the roughness of the pore walls may lead to an over estimation of pore nodes. Small variations in diameter due to the surface roughness can lead to a region being separated into many pores. To avoid this issue, a post-processing procedure is applied to merge neighboring pores that have radius similar to the previously found throat node, such as:
\begin{eqnarray}
(r^i_{throat}/r_{pore}^i)< \lambda for \lambda,0<\lambda<1
\label{eq:merge}.
\end{eqnarray} 
The throats lengths are also defined differently, by a maximum pore length. The MBA defines the border between the end of a pore and the start of a throat as the distance between the pore center and the sphere on the pore-throat chain having a radius of 0.7 times its pore's radius \cite{Dong2009}. The throat length l\textsubscript{t} is defined by subtracting the two pore lengths (l\textsubscript{i} and l\textsubscript{j}) from the total throat length l\textsubscript{i,j} (the distance from pore i center to pore j center).
 
It was chosen not to do the pore merging process since the Pore-Networks were extracted to be compared to the Medial-Axis Networks of the previous step. In addition the behavior of the model for single phase simulations could be verified with less interference of free parameters, which could lead to implementation errors. Moreover, different choices of $\lambda$ between implementations of the algorithm can provide different results on the pore-throat separation for the same sample, which in turn, could lead to different interpretations of the porous structure. 

\begin{figure*}
\includegraphics[scale=0.82]{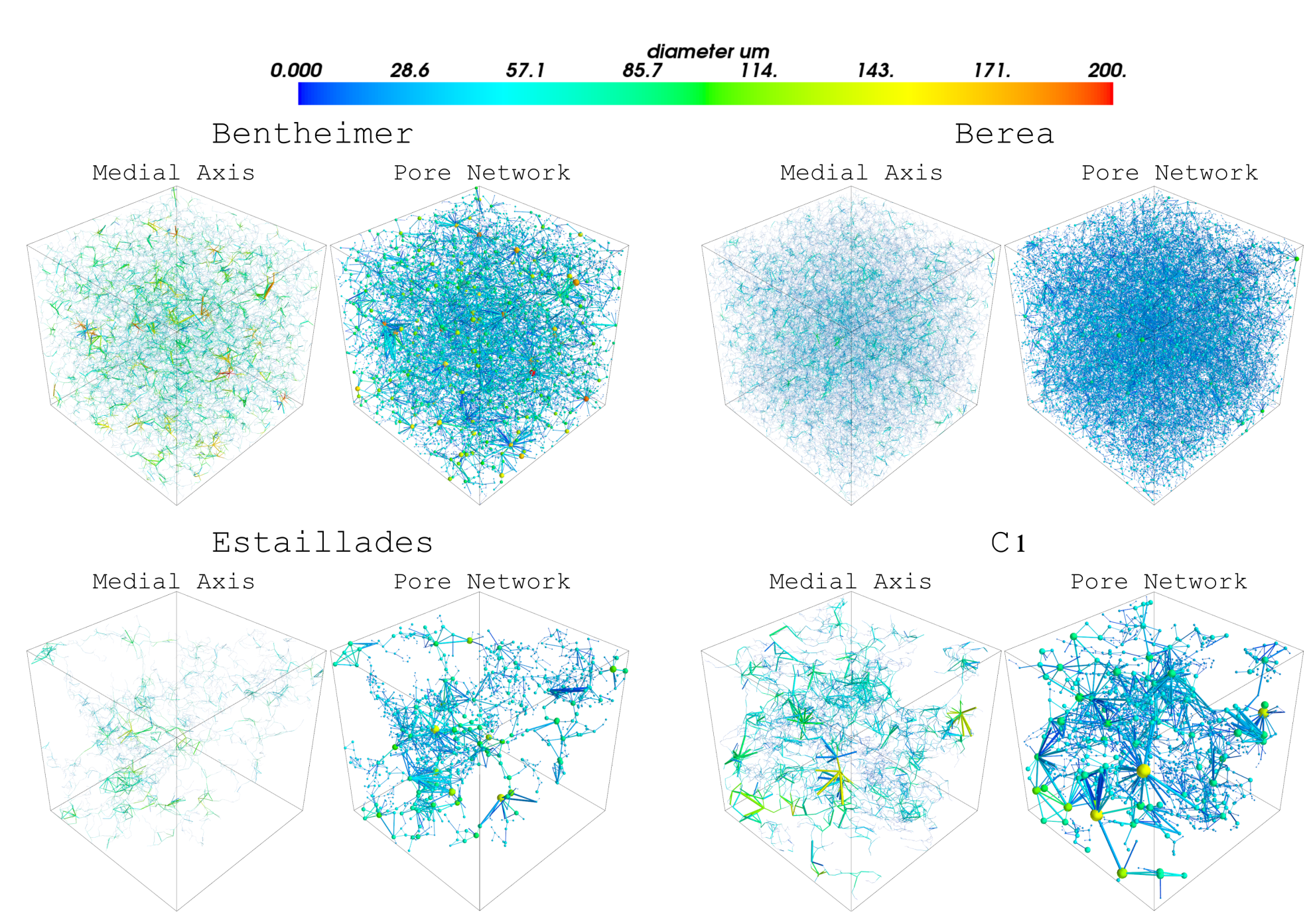}
\caption{\label{fig:samples} The Medial-Axis Networks and Pore Networks obtained by the developed extraction methods}
\end{figure*}

\section{VALIDATION AND DISCUSSIONS}
The method was validated by applying both network extractions to sandstone and carbonate samples presently available on the Imperial College Rock repository \cite{C1,Estaillades,Bentheimer}. The reason for this choice was so the method could be tested by the community on the same set of samples. Another advantage is they are available already in binary form, so possible noise removal steps that could add uncertainty to the results were avoided. And most importantly, the Imperial college repository also provides their extracted networks and simulation results for users to compare. 

\begin{table}
\caption{\label{tab:samples}%
Characteristics of the samples
}
\begin{ruledtabular}
\begin{tabular}{cccc}
\textrm{Name}&
\textrm{Dimensions}&
\textrm{Resolution (\textmu m) }&
\textrm{Porosity}\\
\colrule
Bentheimer & 1000x1000x1000 & 3.0 & 21.6\\
Berea & 1000x1000x1000 & 2.25 & 18.9\\
C1 & 400x400x400 & 2.85 & 23.3\\
Estaillades & 500x500x500 & 3.3 & 12.2\\
\end{tabular}
\end{ruledtabular}
\end{table}

The algorithm was tested on a Estaillades carbonate \cite{Estaillades} sample, a carbonate sample named C1 \cite{C1}, a Bentheimer \cite{Bentheimer} sandstone sample, and finally a Berea sandstone sample. The Berea sample is the only exception for which the tomography images were acquired by our group on a Bruker\textsuperscript{TM} CT Scanner model SkyScan 1272, with 2.25\textmu m voxel resolution. This sample's absolute permeability was obtained experimentally, and it was added to this work so the network's results could be compared directly to experimental data. The characteristics of the samples are summarized on table ~\ref{tab:samples}.

The single-phase simulations were performed using OpenPNM \cite{openpnm} and the morphology evaluation using Porespy \cite{porespy}, both open-source pore network modeling and pore-network extraction tools developed in Python. Wonderful projects that provide well-documented, simple to use and precise tools for pore space network modelling. 

The proposed model's algorithm was developed in Python, then transferred to C++ code for better performance, as the search for the Maximal Balls and shortest paths through Djikstra Algorithm are run in parallel using Threads. The C++ code makes use of the ITK imaging processing package for reading and saving the data and for the distance transform calculations. The graph modeling was performed using the BoostGraph package.The developed algorithm was run on a Dell precision Tower T7910.

\subsection{Medial-Axis Network evaluation}

The networks were first tested for single-phase flow simulations and the estimation of the absolute permeability for the three weighting models. After the connectivity of the pore space is recovered through each model, the hydraulic conductivity for each edge is set as:
\begin{eqnarray}
g_{i,i+1}= (\pi R_{i}^4)/8 \mu L_{i,i+1} 
\label{eq:hyconductivity}.
\end{eqnarray}  
L\textsubscript{i,i+1} is the distance between nodes and R\textsubscript{i} is the radius associated to the edge on the complete network construction step.
The permeability k\textsubscript{d} was calculated through Darcy's law, eq.~\ref{eq:darcy}. The subscript d refers to the orthogonal directions x,y,z of the applied pressure gradient, ${\triangle P_d=1Pa}$. The dynamic viscosity ${\mu}$ was set to 1,0 10\textsuperscript{-3} Pa.s, the dynamic viscosity of water. D\textsubscript{d} is the length of the sample on the direction d and A\textsubscript{d} was set as the effective area of the cross section of the sample (the area of the outlet pores).
\begin{eqnarray}
k_d= (\mu D_d/A_d)Q_d/\triangle P_d
\label{eq:darcy}.
\end{eqnarray} 
The permeability calculated for each model for the four samples is presented on table ~\ref{tab:weights}. The reported permeabilities are either the absolute experimental permeability, or the reported permeabilities presented on the before mentioned references. Although the permeability are usually presented as one absolute value (the experimental value for the Berea and the reported absolute permeability for Bentheimer and Estallaides samples) , it was chosen to show the calculated permeability for the three directions so a further inspection of the method could be made. The exception is the carbonate sample C1, for which the permeability for the three directions is reported: k\textsubscript{x} = 785,k\textsubscript{y} = 1469, k\textsubscript{z} = 1053. The reported value presented for this sample is the average of the results for the three directions.
\begin{table}
\caption{\label{tab:weights}Permeability [mD] calculated for each model.}
\begin{ruledtabular}
\begin{tabular}{ccccccc}
 Sample&Model&kx&ky&kz&Avg&Reported\\ \hline
 \multirow{3}{*}{Bentheimer} &1&6749&7192&7134&7025&\multirow{3}{*}{3547}\\ &2&3689&3872&3891&3817&\\&3&2604&2805&2775&2728& \\ \hline
 \multirow{3}{*}{Berea} &1&1878&1830&2370&2026&\multirow{3}{*}{623}\\ &2&944&907&1187&1012&\\&3&674&881&623&726& \\ \hline 
 \multirow{3}{*}{C1} &1&1970&4178&2617&2921&\multirow{3}{*}{1102}\\ &2&1140&2208&1515&1621&\\&3&868&1508&1075&1150& \\ \hline 
 \multirow{3}{*}{Estaillades} &1&3197&395&441&1344&\multirow{3}{*}{172}\\ &2&441&171&210&274&\\&3&616&49&153&272& \\ \hline
\end{tabular}
\end{ruledtabular}
\end{table}

Model 1, in which no weighting strategy was used, lead to an overestimation of the samples' permeability for all cases. The ratio of the average measured permeability to the reported varied from 7.8 for the Estaillades to 2.65 for C1. It could be argued that the average of the measured permeabilities is not as representative for the Estaillades sample as it is for the others, since there is large variation between the values observed. Nonetheless, the increase of permeability of model 1 for this sample is evident for all directions of measurement. 

The change to Model 2, in which the weights of the edges are set as the cross section area, resulted in average permeabilities that are a close approximation to the reported ones. The overestimation decreased considerably for the Estaillades sample, resulting in 1.6. However, the variations for Berea and C1 were still relevant. 

Model 3 seemed to correct even further the overestimation of the other models. The ratio of the average to the reported values for C1 and Berea were 1.1 and 1.2 respectively. Comparing to the reported values for each direction of C1, k\textsubscript{x} = 785,k\textsubscript{y} = 1469, k\textsubscript{z} = 1053, the values obtained with model 3 were clearly a better approximation, k\textsubscript{x} = 868,k\textsubscript{y} = 1508 and k\textsubscript{z} = 1075.It is interesting to note that for the Bentheimer sample, model 2 had a better approximation to the reported permeability than model 3. This might be due to the characteristic larger pores of the sample ( Fig. ~\ref{fig:samples}).

These results imply that defining the shortest path as the path of least fluid flow resistance (Model 3), ensures that it will follow the medial axis of the object, avoiding redundant information that would lead to a representation of the pore space that is not topological equivalent to the true topology.

The morphology preservation  of the networks of model 3 can be further evaluated by comparing the results from porespy's local thickness algorithm to the volume weighted radius distribution of the nodes preserved on the Medial-Axis extraction.
\begin{figure}
\includegraphics[scale=0.3]{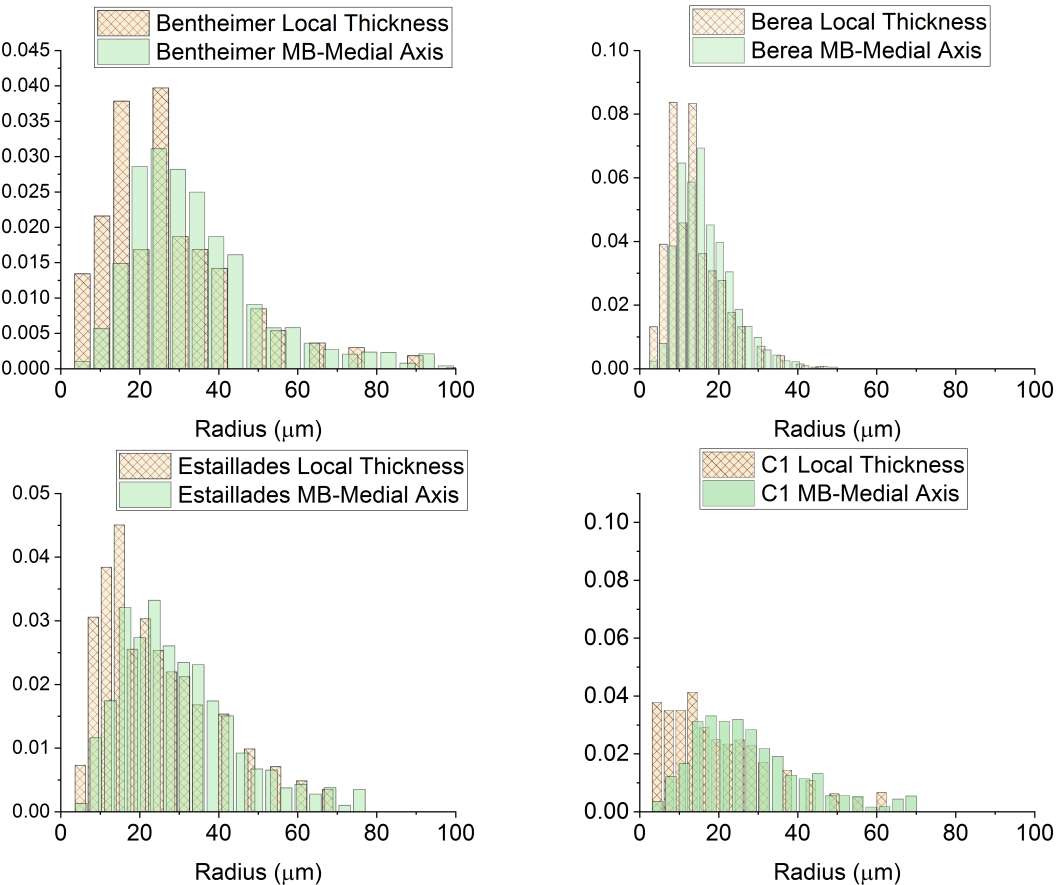}
\caption{\label{fig:local} Morphology Evaluation of the Medial Networks.}
\end{figure}
It is possible to see from the results of Fig.~\ref{fig:local} that the networks are representative of the pore space. The deviations for small radii were expected since corners and fractures are not captured in the medial axis method.

Another question was the influence of the number of pores used as seeds to find the Medial Axis networks. This influence was tested by sorting the Max Balls identified as pore centers by their radius from largest to smallest. Then the networks were generated using model 3, searching for the smallest paths for an increasing number of pore seeds. 
\begin{figure}
\includegraphics[scale=0.4]{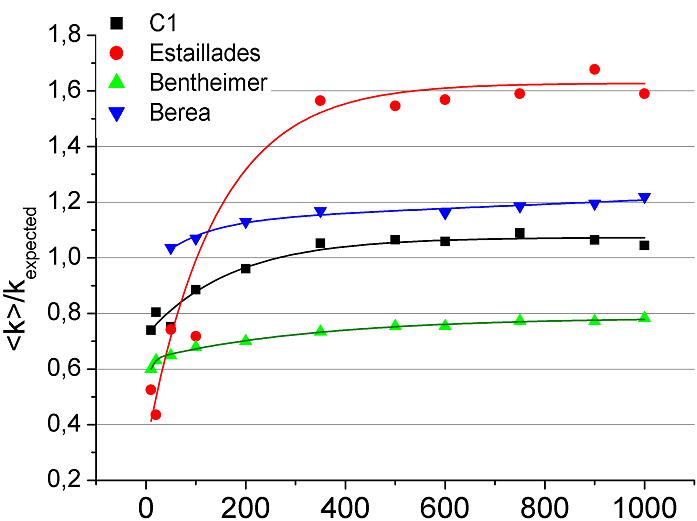}
\caption{\label{fig:porevar} Ratio of the average calculated permeability to the expected (reported or measured) permeability of the samples versus the number of pore spheres used as seeds. The pore spheres are the local maxima sorted from the largest to the smallest sphere.}
\end{figure}
The results were that eventually the average permeability value converges for all samples (Fig.~\ref{fig:porevar}). This also implies that for a large enough number of pore nodes used as seeds, all the medial axis paths are recovered, and further searching is not required. Other iterations only recover the same paths again.

\begin{figure*}
\includegraphics[scale=0.8]{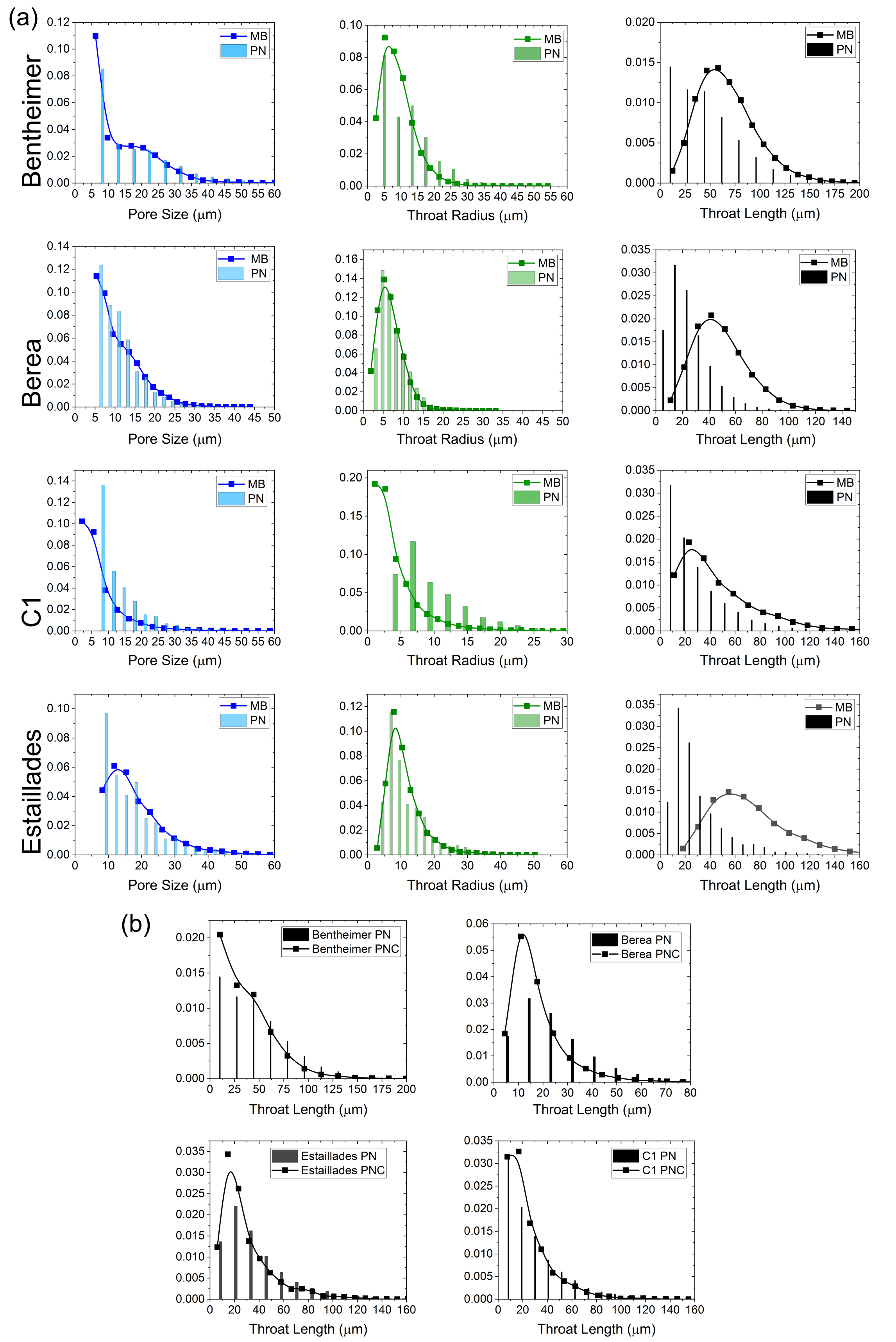}
\caption{\label{fig:poresize}Pore size, Throat Size and Throat Length distributions for the proposed model (PN) in comparison to the original Maximal Ball model (MB). }
\end{figure*}
\subsection{Pore Network evaluation}

Finally, the Medial Axis networks were further processed in order to model the Pore-Throat Networks. Both the Medial-Axis approach as well as the Pore Networks are presented on Fig.~\ref{fig:samples}.

The pore and throat size's acquired by the developed method were compared to the pore and throat size's from the networks extracted by the original Maximal Ball algorithm. The results are presented on Fig.~\ref{fig:poresize}(a), MB is the original Maximal Ball Network, PN is the version extracted using the Medial Axis. We had good agreement in both pore sizes and throat sizes distributions for most samples, even though the proposed method did not apply any restrains on minimum pore size/throat. The exception being the carbonate sample C1, for which the proposed method had the tendency of identifying larger pores and throats.

The major differences between the PN and MB networks were observed on the throat length distributions, Fig.~\ref{fig:poresize}(a). The short throat lengths present on the proposed model is an effect of choosing not to clear the network of small sized pores and not applying the merging technique. This is more evident when we compare the number of pores identified on the PN and MB methods , table ~\ref{tab:topology}. 

\begin{table}
\caption{\label{tab:topology}%
Comparison of the number of pores and coordination numbers between pore networks models. PN refers to the proposed model and MB is the classic Maximal Ball model.
}
\begin{ruledtabular}
\begin{tabular}{cccc}
\textrm{Sample}&
\textrm{Network}&
\textrm{N Pores}&
\textrm{Mean Coord. Number }\\
\colrule
 \multirow{2}{*}{Bentheimer}&MB & 10377 & 4.6\\&PN & 18535 & 2.6\\\hline
 \multirow{2}{*}{Berea}& MB & 22744 & 3.9\\& PN & 51351 & 2.6\\\hline
\multirow{2}{*}{C1}& MB & 2612 & 3.9\\& PN & 3315 & 2.9\\\hline
\multirow{2}{*}{Estaillades}& MB & 2500 & 3.6\\& PN & 2612 & 2.9\\\hline
\end{tabular}
\end{ruledtabular}
\end{table}

The pore space is subdivided into more nodes in the proposed method, leading to more sparse networks, with smaller mean coordination numbers. When MBA merges regions, more distant pores are directly connected, resulting in higher coordination numbers and longer throats. 

However, even though the throats length are short in the proposed approach, the modelling of the total length as the distance between pore centers have had a great impact on the permeability estimations, as it is shown on table ~\ref{tab:pnmedial}. The result gets far more precise with the correction done in the original MBA, represented by PNC (Pore Network Corrected) in Fig. ~\ref{fig:poresize} (b) and on table ~\ref{tab:pnmedial}. 
\begin{table}
\caption{\label{tab:pnmedial}Permeability [mD] of the Pore Networks, without the throat length correction (PN) and corrected version (PNC) in contrast with the calculated permeability of the Medial-Axis Networks.}
\begin{ruledtabular}
\begin{tabular}{ccccc}
 Sample&Network&kx&ky&kz\\ \hline
 \multirow{3}{*}{Bentheimer}&Medial&2916&2604&2805\\&PN&1144&1027&1061\\&PNC&2595&2284&2392\\\hline
 \multirow{3}{*}{Berea}&Medial&722&674&881\\& PN&301&283&376\\& PNC&505&468&623\\\hline
 \multirow{3}{*}{C1}& Medial&868&1508&1075\\& PN&385&668&473\\& PNC&802&1164&994\\\hline
 \multirow{3}{*}{Estaillades}& Medial&616&49&153\\& PN&258&12&61\\& PNC&426&24&133\\\hline
\end{tabular}
\end{ruledtabular}
\end{table}
This effect was surprising, since the distances between pores were already small. The few longer throats lead to an underestimation of the permeability by a factor of almost 2 in most cases. Although only a subtle change can be seen on ~\ref{fig:poresize} (b), it had a major impact on the final results.

Finally, in terms of performance, the current implementation performed well in comparison to the MBA. The fastest case was for sample C1, in which the distance transform,  Maximal Balls  and Medial Axis network is found after a few minutes. The processing for the Estaillades and Bentheimer samples was complete in 4 minutes for 20 seeds and 20 minutes for 1000 seeds. The slowest case was for the Berea sample, which the total processing time took a little more than 60 minutes using 1000 pores nodes as pore seeds. 

The reason for the slower performance for the Berea sample could be due to the high porosity, sample size and pore sizes. Even though the total porosity is similar, the larger pore sizes found in the Bentheimer sample lead to the complete Maximal Balls network with fewer nodes, the reason why the algorithm performance for this sample was better than for the Berea sample.

The memory usage has reached as far 50GB for the Berea sample. However, the algorithm is not completely optimized memory wise. The process uses copies of the complete matrix representing the total volume on the graph modelling step, which could be avoided in the future. The processing to acquire the Pore-throat Networks and single phase simulations were done in a Python script, with good performance for all samples. 

\section{CONCLUSIONS}
In this work, a new method for the extraction of a Medial Axis Network from the basis of the Maximal Ball Algorithm was presented. 

The model, which transforms the Maximal Balls into a large network and then searches for the optimum paths, was tested for three different weighting strategies. It was shown that the network with no weights assigned lead to Medial-Axis Networks that overestimate the samples' absolute permeabilities. The networks found by both weighting strategies of inscribed sphere area and Hagen-Pouseille models gave better results for single phase flow simulations. The comparison to the reported permeabilities showed that the Hagen-Pouseille based method was the best suited model for these samples. The morphology of the Medial-Axis networks were further investigated by comparing the diameter variation of the spheres to a local thickness measurement tool, with good agreement for all cases.

Afterwards, the Pore Networks extracted using the results of the Medial-Axis method were compared to the networks acquired through the original Maximal Ball Algorithm. The pore size and throat size distributions are very similar between the methods, however the throat lengths are much smaller in the Pore Networks extracted in comparison to the original Max Ball Algorithm. After applying the correction proposed originally in the MBA, the absolute permeability estimation were a fair approximation to the expected results. This evaluation of the method is an important insight of the impact of the chosen parameters for the network extraction, which should be further investigated in future works.

\begin{acknowledgments}
The group would like to thank Imperial College for making available the samples for testing new methods for porous media analysis. We also would like to thank Dr. Rodrigo Neumann and Professor Dr. Francisco Aparecido Rodrigues for the insights and all the help. This study was financed in part by the following Brazilian institutions: Coordenação de Aperfeiçoamento de Pessoal de Nível Superior - Brasil (CAPES) - Finance Code 001, National Council for Scientific and Technological Development (CNPq) (153627/2012-3,140215/2015-8), University of São Paulo (USP), and Centro de Pesquisa e Desenvolvimento Leopoldo Américo Miguez de Mello (Cenpes/Petrobras) (ANP 2014/00389-8,2015/00416-8).  
\end{acknowledgments}


\bibliography{apssamp.bib}

\end{document}